\documentclass{iaus}
\usepackage{graphicx}
\usepackage{amsmath,amssymb}
\usepackage{graphicx}
\usepackage{natbib}
\usepackage[usenames]{color}
\usepackage{ifpdf}

\newcommand{\beq}{
\begin{equation}
}
\newcommand{\eeq}{
\end{equation}
}
\newcommand{\beqa}{
\begin{eqnarray}
}
\newcommand{\eeqa}{
\end{eqnarray}
}

\newcommand{\sigmaconf}   {\ensuremath{\sigma_{68}}}
\newcommand{\msint} {\ensuremath{8.12}}
\newcommand{\msinterr} {\ensuremath{8.12 \pm 0.08}}
\newcommand{\msslope} {\ensuremath{4.24}}
\newcommand{\msslopeerr} {\ensuremath{4.24 \pm 0.41}}

\newcommand{\msscaterr} {\ensuremath{0.44 \pm 0.06}}

\newcommand{\mlint} {\ensuremath{8.95}}
\newcommand{\mlinterr} {\ensuremath{8.95 \pm 0.11}}
\newcommand{\mlslope} {\ensuremath{1.11}}
\newcommand{\mlslopeerr} {\ensuremath{1.11 \pm 0.18}}

\newcommand{\mlscaterr} {\ensuremath{0.38 \pm 0.09}}

\makeatletter
\newcommand\farcm@mss{\mbox{$.\mkern-4mu^\prime$}}%
\let\farcm\farcm@mss
\newcommand\farcs@mss{\mbox{$.\!\!^{\prime\prime}$}}%
\let\farcs\farcs@mss
\makeatother

\title[Scatter in $M$--$\sigma$ and $M$--$L$]
{Determination of the intrinsic scatter in the $M$--$\sigma$ and
$M$--$L$ relations}

\author[Kayhan G\"ultekin]   
{Kayhan G\"ultekin$^1$}

\affiliation{$^1$Department of Astronomy, University of Michigan,
\\500 Church Street, Ann Arbor, MI, 48109, USA \\ E-mail: {\tt
kayhan@
umich.edu}}

\pubyear{2009}
\volume{267}  
\pagerange{119--126}
\jname{Co-Evolution of Central Black Holes and Galaxies}
\editors{B.M.\ Peterson, R.S.\ Somerville, \& T.\ Storchi-Bergmann, eds.}

\newcommand{\msun}     {\ensuremath{{M}_{\scriptscriptstyle \odot}}}
\newcommand{\kms}      {\ensuremath{\,\mathrm{km~s^{-1}}}}

\newcommand{\msigma}   {\ensuremath{M}{--}\ensuremath{\sigma}}
\newcommand{\ml}       {\ensuremath{M}{--}\ensuremath{L}}
\newcommand{\mbh}      {\ensuremath{M_{\mathrm{BH}}}}

\newcommand{\rinfres} {\ensuremath{R_{\mathrm{infl}} / r_{\mathrm{res}}}}
\newcommand{\rinfl} {\ensuremath{R_{\mathrm{infl}}}}

\begin{document}

\maketitle

\begin{abstract}
We derive improved versions of the relations between supermassive
black hole mass ($M_\mathrm{BH}$) and host-galaxy bulge velocity
dispersion ($\sigma$) and luminosity ($L$) (the \msigma\ and \ml\
relations), based on $\sim50$ \mbh\ measurements and $\sim20$ upper
limits.  Particular attention is paid to recovery of the intrinsic
scatter ($\epsilon_0$) in both relations.  We find the scatter to be
significantly larger than estimated in most previous studies.  The
large scatter requires revision of the local black hole mass function,
and it implies that there may be substantial selection bias in studies
of the evolution of the \msigma\ and \ml\ relations.  When only
considering ellipticals, the scatter decreases.  These results appear
to be insensitive to a wide range of assumptions about the measurement
errors and the distribution of intrinsic scatter.  We also investigate
the effects on the fits of culling the sample according to the
resolution of the black hole's sphere of influence.
\end{abstract}

\firstsection 
\section{Overview}

The \ml\ and \msigma\ relations --- the relations between a black
hole's mass and the host galaxy's (bulge) luminosity, $L$, or velocity
dispersion, $\sigma$ ---
strongly suggest a fundamental link between galaxy and black hole (BH)
evolution \citep{dressler89,kormendy93a,magorrianetal98,gebhardtetal00a,fm00}.  In this contribution to the proceedings, we discuss recent
developments in the study of these relations and how they relate to
coevolution of black holes and galaxies.
Fundamental to the understanding of the \msigma\ and \ml\ relations is the
measurement of the relation's \emph{intrinsic} or \emph{cosmic
scatter}, as distinct from scatter due to measurement errors.  The
fact that there is a relation between BH mass and stellar velocity
dispersion is not surprising, but the scatter is remarkably small,
estimated by \citet{tremaineetal02} to be no larger than 0.25--0.3
dex.  The \emph{total} scatter of the relations (intrinsic combined
with statistical and systematic measurement errors) is what makes the
relation useful as a secondary tool for BH mass estimation.  The
intrinsic scatter, however, is the measure of the fundamental link between the
physical quantities in question.

The magnitude of the intrinsic scatter is extremely important for
several reasons.  First, the range of BH masses in galaxies of a given
velocity dispersion or bulge luminosity constrains BH formation and
evolution theories.  Many theories of BH formation and galaxy
evolution have used the \msigma\ relation either as a starting point
for further work or as a prediction of the theory
\citep[e.g.,][]{sr98,1999MNRAS.308L..39F,2003ApJ...596L..27K}; for a
review, see \citet{richstone04}.  A further test of such theories is
whether they can reproduce the observed cosmic scatter in the
relation.  For example, there may be an increased intrinsic scatter in
low-mass galaxies because BHs are ejected by asymmetric gravitational
wave emission and low-mass spheroids have lower escape
velocities \citet{volonteri07,vhg08}.

Understanding the scatter in the \msigma\ relation is also essential
for estimating the space density of the most massive BHs in the local
universe.  One of the most useful aspects of the \msigma\ relation is
that it allows one to estimate a galaxy's central BH mass from the
more easily measured velocity dispersion.  Because of the steep
decline in number density of galaxies having high velocity dispersion
\citep{shethetal03, bernardietal06,laueretal07}, the majority of the
extremely large BHs will reside in galaxies with moderate velocity
dispersions that happen to contain BHs that are overmassive for the
given velocity dispersion \citep{yt02,marconietal04,laueretal07}.
Knowing the magnitude of the intrinsic scatter is thus required to
find the density of the most massive BHs.  For example, the number
density of BHs with $M > 10^{10}~\msun$ is $\sim 3~\mathrm{Gpc}^{-3}$
if the intrinsic scatter is 0.15 dex and $\sim 30~\mathrm{Gpc}^{-3}$
if the intrinsic scatter is 0.30 dex \citep{laueretal07}.

Both the magnitude of the intrinsic scatter and its distribution
(e.g., normal or log-normal in mass) are also important to know for
studies of the evolution of the \msigma\ relation
\citep[e.g.,][]{tmb04, treuetal07, hopkinsetal06, pengetal06,
shenetal07, shenetal08, vestergaardetal08}.  \citet{laueretal07c}
showed that there is a bias when comparing BH masses derived from
observations of inactive galaxies at low redshifts to BH masses from
active galaxies at higher redshift.  The bias arises because the
sample of nearby galaxies measures the distribution of BH masses for a
given host velocity dispersion or luminosity, whereas the sample from
high-redshift galaxies tends to measure the distribution of the host
luminosity or host velocity dispersion for a given BH mass.
\citet{laueretal07c} found that the bias in the inferred logarithmic
mass scales as the square of the intrinsic scatter in logarithmic
mass.  In order to account for this bias correctly, not only the
magnitude but also the distribution of the deviations from the
\msigma\ relation is needed.

\section{Most Recent Scaling Relations and their Scatter}

\begin{figure}[htb]
\centering
\includegraphics[width=0.80\textwidth]{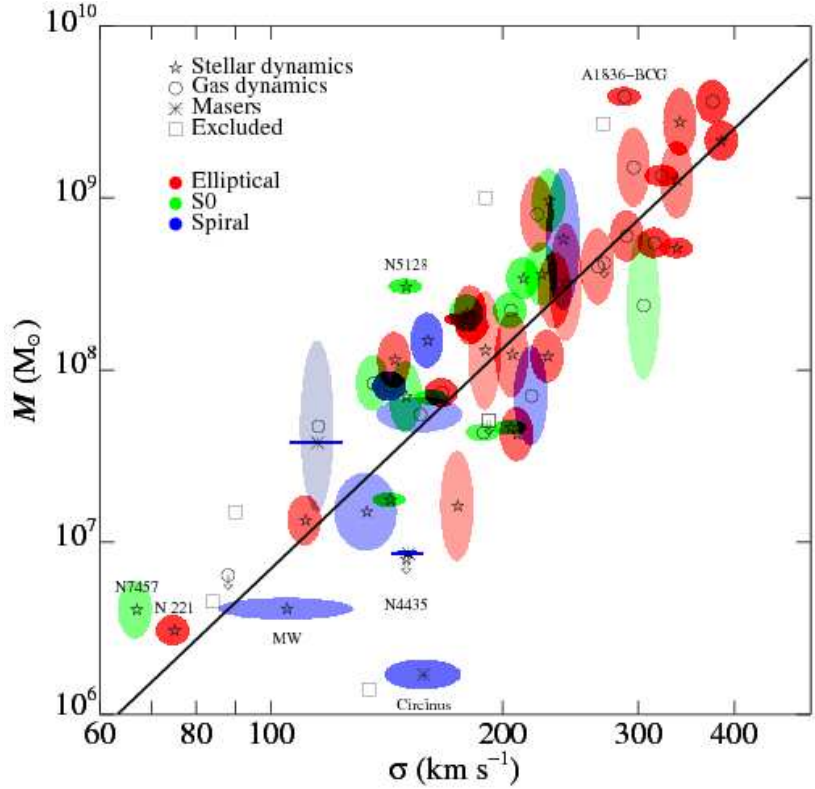}
\caption{The \msigma\ relation for galaxies with dynamical
measurements.  The symbol indicates the method of BH mass measurement:
stellar dynamical (\emph{pentagrams}), gas dynamical (\emph{circles)},
masers (\emph{asterisks}).  Arrows indicate 3$\sigmaconf$ upper limits
to BH mass.  The shade of the error ellipse indicates the Hubble type
of the host.  The saturation of the shades in the error ellipses is
inversely proportional to the area of the ellipse.  The line is
the best fit relation to the full sample: $M_{\mathrm{BH}} =
10^{\msint}~\msun(\sigma/200~\kms)^{\msslope}$.  The mass uncertainty
for NGC~4258 has been plotted much larger than its actual value so
that it will show on this plot.  (Adapted from
\citealt{gultekinetal09b}.)}
\label{f:msigma}
\end{figure}

\begin{figure}[htb]
\centering
\includegraphics[width=0.80\textwidth]{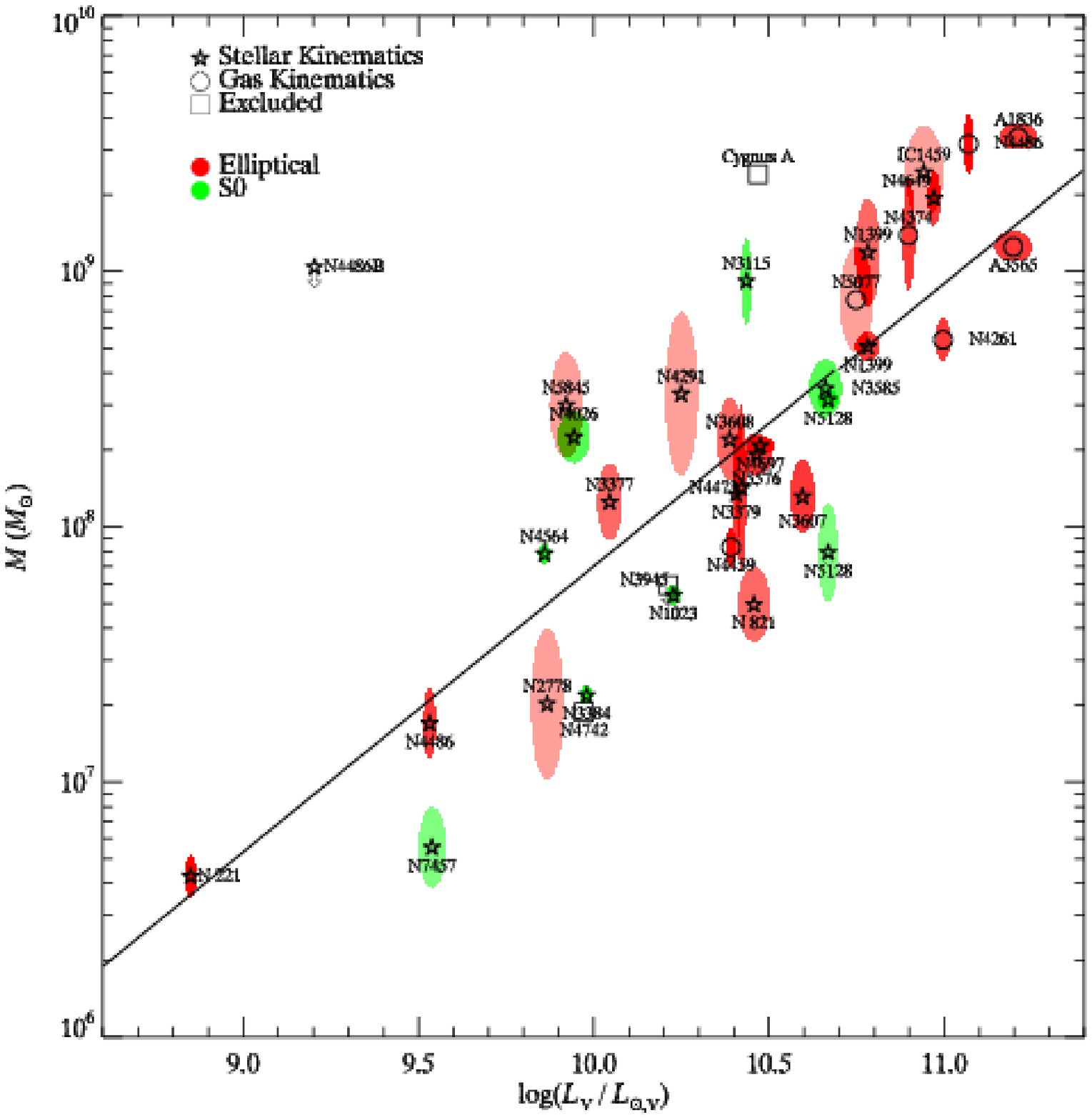}
\caption{The $M$--$L$ relation for galaxies with dynamical
measurements.  The symbol indicates the method of BH mass measurement:
stellar dynamical (\emph{pentagrams}) and gas dynamical
(\emph{circles)}.  Arrows indicate upper limits for BH mass.  The
shade of the error ellipse indicates the Hubble type of the host
galaxy, and the saturation of the shade is inversely proportional to
the area of the ellipse.  The line is the best-fit relation for the
sample without upper limits: $M_{\mathrm{BH}} = 10^{\mlint}~\msun
(L_V/10^{11}~{\rm L}_{{\scriptscriptstyle \odot},V})^{\mlslope}$.
(Adapted from \citealt{gultekinetal09b}.)}
\label{f:ml}
\end{figure}

Using a sample of $\sim50$ BH mass measurements and $\sim20$,
\citet{gultekinetal09b} used a generalized maximum likelihood method
to fit the \msigma\ and \ml\ relations with an intrinsic scatter
component.  Figure~\ref{f:msigma} and figure~\ref{f:ml} show \msigma\ and \ml\
relations.  Their best fit for \msigma\ was
\beq
\log{\left(\frac{M_{\mathrm{BH}}}{\msun}\right)} = \msinterr + \left(\msslopeerr\right)\log{\left(\frac{\sigma}{200~\kms}\right)}
\eeq
with an intrinsic scatter distributed normally in logarithmic mass
with standard deviation $\epsilon_0 = \msscaterr$.  When considering
only ellipticals, they found that the intrinsic scatter drops to
$\epsilon_0 = 0.31 \pm 0.06$.  Their best fit for \msigma\ was
\beq
\log{\left(\frac{M_{\mathrm{BH}}}{\msun}\right)} = \mlinterr + \left(\mlslopeerr\right)\log{\left(\frac{L_V}{L_{\odot, V}}\right)}
\eeq
with an intrinsic scatter distributed normally in logarithmic mass
with standard deviation $\epsilon_0 = \mlscaterr$.  Different
assumptions about the error distribution in black hole mass
measurements and intrinsic scatter did not significantly alter these
conclusions.

\section{Implications for Scatter: Potential for Bias in Co-evolution Studies}

The intrinsic scatter is essential for determining the cosmic density
of the most massive black holes.  Because of the exponential drop in
number density at the high end, very large galaxies are extremely
rare.  This means that in a fixed volume, the greatest number of big
black holes do not come from the intrinsically-rare, large galaxies
but from the more common modest-sized galaxies that happen to have an
over-massive black hole.  Thus, deriving the number density of the
largest black holes from the number density of galaxies depends on
this effect.

The tendency for the largest black holes to come from more modestly
sized galaxies also leads to a potential bias in studies of the
evolution of scaling relations \citep{laueretal07c}.  The samples of
high-redshift black holes, which are measured from AGNs, tend to probe
distribution of $\sigma$ or $L$ for a given black hole mass whereas
the local, quiescent sample tends to probe the distribution of \mbh\ 
for a given host galaxy property.  Direct comparison of the two
samples leads to a bias that would lead one to incorrectly infer that
at high redshift black holes were more massive than they are at low
redshift, for a given velocity dispersion or bulge luminosity.  This
bias depends critically on the poorly constrained wings of the
distribution of the intrinsic scatter.

Another, previously neglected caveat is that it may be impossible to
disentangle evolution of just the intrinsic scatter of the relations
from the evolution of the slope and/or intercept of the relations.
That is, to correct for this effect when looking for evolution of
\msigma\ or \ml, one needs to assume that at high redshift the
intrinsic scatter is the same as at low redshift in the quiescent
population.  This assumption that one aspect of the relation (the
scatter) is constant while measuring changes in other aspects (the
intercept and slope) is difficult to justify and may lead to erroneous
conclusions.  Finally, a very recent study \citep{2009arXiv0911.5208S}
has shown that an independent bias arising from uncertainties in BH
mass estimators and the shape of the BH mass function may lead to an
overestimate of the true BH masses.

\section{Scaling Relations are Biased When Black Hole Masses are Censored by Sphere-of-Influence Resolution }

\begin{figure}
\begin{center}
\includegraphics[width=0.55\textwidth,angle=90]{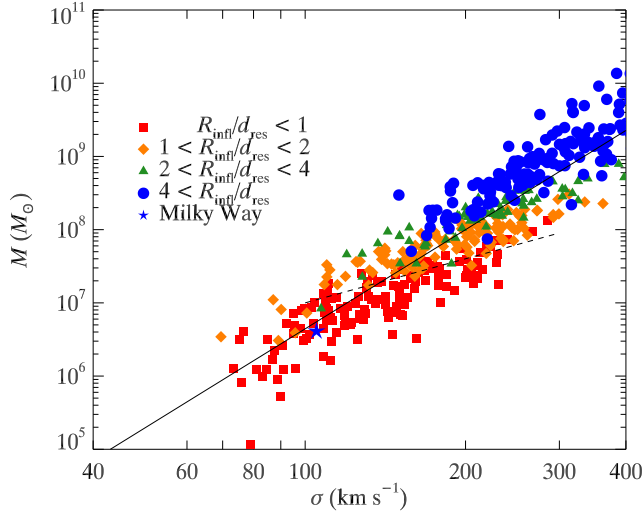}
\caption{This figure shows results of a synthetically generated sample
of 500 galaxies with BH mass generated from an \msigma\ relation with
$\alpha = 8$ and $\beta = 4.0$ and a log-normal scatter of 0.3 dex
with measurement errors of 0.2 dex.  The \msigma\ ridge line is drawn
as a black line.  The Galaxy is plotted as a pentagram.  Different
symbols indicate different levels of resolution: $\rinfres < 1.0$
(\emph{squares}), $1.0 < \rinfres < 2.0$ (\emph{diamonds}), $2.0 <
\rinfres < 4.0$ (\emph{triangles}), and $\rinfres > 4.0$
(\emph{circles}).  Fitting the combined subsamples with \rinfres\
exceeding a given value yields biased estimators compared to the
underlying \msigma\ relation..  The reason for the bias in slope is
that cuts in $R_{\mathrm{infl}}$ tend to fall along lines of $\mbh
\propto \sigma^{\beta - 2}$ (since $R_{\mathrm{infl}} \propto \mbh
\sigma^{-2}$ and $\mbh \propto \sigma^\beta$).  This is illustrated by
the dashed line of slope 2.0.  (Adapted from
\citealt{gultekinetal09b}.)}
\end{center}
\label{f:rinfmc}
\end{figure}

Some have advocated the removal of BHs from scaling relation
estimation if the BH's radius of influence on the sky is below a given
threshold \citep[e.g.,][]{ff05}.  This is based on a misconception
that under-resolved BHs will yield biased BH mass estimates.  In fact,
under-resolved BHs yield lower precision masses, but with no bias in
accuracy \citep{gultekinetal09a,gultekinetal09b}.  On the other hand,
we may demonstrate the effects of censoring data based on \rinfl with
a synthetic \msigma\ data set.  The synthetic data set consists of a
sample of 500 galaxies, uniformly distributed in volume out to a
distance of 30~Mpc.  Each galaxy is given a velocity dispersion from a
normal distribution in $\log{(\sigma/200~\kms)}$ centered at 0 with
standard deviation 0.2.  Each galaxy is given a BH mass from an
\msigma\ relation with intercept $\alpha = 8$, slope $\beta = 4.0$,
and log-normal intrinsic scatter with $\epsilon_0 = 0.3$~dex.  The
BH's logarithmic mass is measured with a normally distributed
measurement error of 0.2~dex and the velocity dispersion has a 5\%
error.  Since each galaxy has a distance, a BH mass, and a velocity
dispersion, we calculate $R_{\mathrm{infl}} = G \mbh \sigma^{-2}$ and
assume the galaxy to be observed with an instrument with resolution of
$d_{\mathrm{res}} = 0\farcs1$.  The results (Fig.~\ref{f:rinfmc}) show
that cutting out BH masses based on their level of resolution would
clearly skew fits to the data set since $\rinfl \sim \mbh \sigma^{-2}$
and $\mbh \sim \sigma^{\beta}$ so that $\rinfl \sim \mbh \sigma^{\beta
- 2}$, resulting in biasing cuts across the data set
\citep{gultekinetal09b}.

\section{Future Work on Intrinsic Scatter}

The current data set of BH masses is insufficient to test some
important questions about scaling relations, especially in regards to
their intrinsic scatter.  One of the most important is whether the
magnitude of the intrinsic scatter changes across galaxy size.  This
would change the extent of the \citet{laueretal07c} bias.  In order to
accurately measure the intrinsic scatter across galaxy size, the
existing data must be augmented.

\bibliographystyle{mn2e}
\bibliography{gultekin}

\end{document}